\documentclass[aps,prl,twocolumn,showpacs,superscriptaddress,groupedaddress]{revtex4}

\usepackage{graphicx}
\usepackage{dcolumn}
\usepackage{bm}
\usepackage{amssymb}
\usepackage{soul}
\usepackage{color}

\def\be{\begin{equation}}
\def\ee{\end{equation}}
\def\bea{\begin{eqnarray}}
\def\eea{\end{eqnarray}}

\begin{document}

\title{Transport measurements of the spin wave gap of thin Mn films}
\author{S.~Buvaev} \affiliation{University of Florida, Gainesville, Florida, 32611, USA}
\author{S.~Ghosh} \affiliation{University of Florida, Gainesville, Florida, 32611, USA}
\author{K.~Muttalib} \affiliation{University of Florida, Gainesville, Florida, 32611, USA}
\author{P.~W\"olfle} \affiliation{Institute for Theory of Condensed Matter and Institute for Nanotechnology, KIT, 76021 Karlsruhe, Germany}
\author{A.~Hebard} \email[Corresponding author:~]{afh@phys.ufl.edu} \affiliation{University of Florida, Gainesville, Florida, 32611, USA} 

\begin{abstract}
Temperature dependent transport measurements on ultrathin antiferromagnetic
Mn films reveal a heretofore unknown non-universal weak localization
correction to the conductivity which extends to disorder strengths greater than
100~k$\Omega$ per square. The inelastic scattering of electrons off of
gapped antiferromagnetic spin waves gives rise to an inelastic scattering
length which is short enough to place the system in the 3d regime. The
extracted fitting parameters provide estimates of the energy gap ($\Delta
\approx$~16~K) and exchange energy ($\bar{J} \approx$~320~K). 
\end{abstract}

\maketitle
\date{\today}

Thin-film transition metal ferromagnets (Fe, Co, Ni, Gd) and
antiferromagnets (Mn, Cr) and their alloys are not only ubiquitous in
present day technologies but are also expected to play an important role in
future developments~\cite{thompson_2008}. Understanding magnetism in these
materials, especially when the films are thin enough so that disorder plays
an important role, is complicated by the long standing controversy about the
relative importance of itinerant and local moments~\cite%
{slater_1936,van_vleck_1953,aharoni_2000}. For the itinerant transition
metal magnets, a related fundamental issue centers on the question of how
itinerancy is compromised by disorder. Clearly with sufficient disorder the
charge carriers become localized, but questions arise as to what happens to
the spins and associated spin waves and whether the outcome depends on the
ferro/antiferro alignment of spins in the itinerant parent. Ferromagnets
which have magnetization as the order parameter are fundamentally different
than antiferromagnets which have staggered magnetization (i.e., difference
between the magnetization on each sublattice) as the order parameter~\cite%
{blundell_2001}. Ferromagnetism thus distinguishes itself by having soft
modes at zero wave number whereas antiferromagnets have soft modes at finite
wave number~\cite{belitz_2005}. Accordingly, the respective spin wave
spectrums are radically different. These distinctions are particularly
important when comparing quantum corrections to the conductivity near
quantum critical points for ferromagnets~\cite{paul_2005} and
antiferromagnets~\cite{syzranov_2012}.

Surprisingly, although there have been systematic studies of the effect of
disorder on the longitudinal $\sigma_{xx}$ and transverse $\sigma_{xy}$
conductivity of ferromagnetic films~\cite%
{bergmann_1978,bergmann_1991,mitra_2007,misra_2009,kurzweil_2009}, there
have been few if any such studies on antiferromagnetic films. In this paper
we remedy this situation by presenting transport data on systematically
disordered Mn films that are sputter deposited in a custom designed vacuum
chamber and then transferred without exposure to air into an adjacent
cryostat for transport studies to low temperature. The experimental
procedures are similar to those reported previously: disorder, characterized
by the sheet resistance $R_0$ measured at $T=$~5~K, can be changed either by
growing separate samples or by gentle annealing of a given sample through
incremental stages of disorder~\cite{misra_2011}. Using these same procedures our results for
antiferromagnets however are decidedly different. The data are well
described over a large range of disorder strengths by a non-universal three
dimensional (3d) quantum correction that applies only to spin wave gapped
antiferromagnets. This finding implies the presence of strong inelastic
electron scattering off of antiferromagnetic spin waves. The theory is
validated not only by good fits to the data but also by extraction from the
fitting parameters of a value for the spin wave gap $\Delta$ that is in
agreement with the value expected for Mn. On the other hand, the
exchange energy $\bar{J}$ could be sensitive to the high disorder in our
ultra thin films, and it turns out to be much smaller compared to the known values.

In previous work the inelastic scattering of electrons off of spin waves has
been an essential ingredient in understanding disordered ferromagnets. For
example, to explain the occurrence of weak-localization corrections to the
anomalous Hall effect in polycrystalline Fe films~\cite{mitra_2007}, it was
necessary to invoke a contribution to the inelastic phase breaking rate $%
\tau_{\varphi}^{-1}$ due to spin-conserving inelastic scattering off
spin-wave excitations. This phase breaking rate, anticipated by theory~\cite%
{tatara_2004} and seen experimentally in spin polarized electron energy loss
spectroscopy (SPEELS) measurements of ultrathin Fe films~\cite%
{plihal_1999,zhang_2010}, is linear in temperature and significantly larger
than the phase breaking rate due to electron-electron interactions, thus
allowing a wide temperature range to observe weak localization corrections~%
\cite{mitra_2007}. The effect of a high $\tau_{\varphi}^{-1}$ due to
inelastic scattering off spin-wave excitations is also seen in Gd films
where in addition to a localizing log($T$) quantum correction to the
conductance, a localizing linear-in-$T$ quantum correction is present and is
interpreted as a spin-wave mediated Altshuler-Aronov type correction to the
conductivity~\cite{misra_2009}.

Interestingly, this high rate of inelastic spin rate scattering becomes even
more important for the thinnest films as shown in theoretical calculations
on Fe and Ni which point to extremely short spin-dependent inelastic mean
free paths~\cite{hong_2000} and in spin-polarized electron energy-loss
spectroscopy (SPEELS) measurements on few monolayer-thick Fe/W(110) films in
which a strong nonmonotonic enhancement of localized spin wave energies is
found on the thinnest films~\cite{zhang_2010}.

Inelastic spin wave scattering in highly disordered ferromagnetic films can
be strong enough to assure that the associated $T$-dependent dephasing
length $L_{\varphi }(T)=\sqrt{D\tau _{\varphi }}$ (with $D$ the diffusion
constant)~\cite{lee_1985} is less than the film thickness $t$, thus putting
thin films into the 3d limit where a metal-insulator
transition is observed~\cite{misra_2011}. Recognizing that similarly high
inelastic scattering rates must apply to highly disordered antiferromagnetic
films, we first proceed with a theoretical approach that takes into account
the scattering of antiferromagnetic spin waves on the phase relaxation rate
and find a heretofore unrecognized non-universal 3d weak localization
correction to the conductivity that allows an interpretation of our experimental
results. 

We mention in passing that the 3d interaction-induced quantum correction 
found to be dominant in the case of ferromagnetic Gd 
films, which undergo a metal-insulator transition\cite{misra_2011}, is
found to be much smaller in the present case and will not be considered further (for an estimate of this contribution see \cite{muttalib_unpub}).

As discussed in detail in Ref.~[\onlinecite{wm10}], the phase relaxation
time $\tau _{\varphi }$ limits the phase coherence in a particle-particle
diffusion propagator $C(q,\omega )$ (Cooperon) in the form 
\begin{equation}
C(q,\omega _{l})=\frac{1}{2\pi N_{0}\tau ^{2}}\frac{1}{Dq^{2}+|\omega
_{n}|+1/\tau _{\varphi }}
\end{equation}
where $N_{0}$ is the density of states at the Fermi level, $\tau $ is the
elastic scattering time and $\omega _{n}=2\pi nT$ is the Matsubara
frequency. Labeling the Cooperon propagator in the absence of interactions
as $C_{0}$, we can write 
\begin{equation}
\frac{1}{\tau _{\varphi }}=\frac{1}{2\pi N_{0}\tau ^{2}}[C^{-1}-C_{0}^{-1}].
\end{equation}
In general, $C(q,\omega )$ can be evaluated diagrammatically in the presence
of interactions and disorder in a ladder approximation \cite{fa} that can be
symbolically written as $C=C_{0}+C_{0}KC$ where the interaction vertex $K$
contains self energy as well as vertex corrections due to both interactions
and disorder. It then follows that $1/\tau _{\varphi }$ is given by 
\begin{equation}
\frac{1}{\tau _{\varphi }}=-\frac{1}{2\pi N_{0}\tau ^{2}}K.
\end{equation}%

In Ref.~[\onlinecite{wm10}], the leading temperature and disorder dependence
of the inelastic diffusion propagator was evaluated diagrammatically, in the
presence of ferromagnetic spin-wave mediated electron-electron interactions.
Here we consider the antiferromagnetic case. We only consider large
spin-wave gap where the damping can be ignored. Using the antiferromagnetic
dispersion relation $\omega _{q}=\Delta +A_{s}q$, where $A_{s}$ is the spin
stiffness, the inelastic lifetime is given by 
\be
\frac{\hbar }{\tau _{\varphi }}=\frac{4 \bar{J}^2}{\pi \hbar n}\int_{0}^{1/l}%
\frac{q^{d-1}dq}{\sinh \beta \omega _{q}}\frac{Dq^{2}+1/\tau _{\varphi }}{%
(Dq^{2}+1/\tau _{\varphi })^{2}+\omega _{q}^{2}}
\ee%
where $n=k_{F}^{3}/3\pi ^{2}$ is the 3d carrier density, $\bar J$ is an effective
spin-exchange interaction and $\beta =1/k_{B}T$. Here we will consider the
limit $\hbar /\tau _{\varphi }\ll \Delta $, relevant for our experiment on
Mn. In this limit we can neglect the $1/\tau _{\varphi }$ terms inside the
integral. The upper limit should be restricted to $\Delta /A_{s}$ in the limit $%
\Delta /A_{s}<1/l$ where $l$ is the elastic mean free path. For large disorder, we expect the parameter $x\equiv 
\hbar Dk_{F}^{2}\Delta / \bar{J}^{2}\ll 1$, where the spin-exchange energy
is related to the spin stiffness by $\bar{J}=A_{s}k_{F}$. In this limit, $L_{\varphi }$ can be
simplified as 
\be
k_{F}L_{\varphi }\approx \left( \frac{\bar{J}}{\Delta }\right) ^{3/2}\left( 
\frac{5\sinh \frac{\Delta }{T}}{12\pi }\right) ^{1/2},\;\;\;x\ll 1
\label{L-phi-3d}
\ee%
which is independent of $x$, and therefore, independent of disorder.

Given the inelastic lifetime, the weak localization correction in 3d is
usually given by \cite{lee_1985} $\delta \sigma _{3d}=e^{2}/\hbar \pi
^{3} L_{\varphi },$ where the prefactor to the inverse inelastic
length is a universal number, independent of disorder. However, at large
enough disorder, we show that there exists a disorder dependent correction,
due to the scale dependent diffusion coefficient near the Anderson
metal-insulator transition. In fact, the diffusion coefficient obeys the
self consistent equation \cite{WV} 
\begin{equation}
\frac{D_{0}}{D(\omega )}=1+\frac{k_{F}^{2-d}}{\pi m}\int_{0}^{1/l}dQ\frac{%
Q^{d-1}}{-i\omega +D(\omega )Q^{2}}
\end{equation}%
where $D_{0}=v_{F}l/d$ is the diffusion coefficient at weak disorder. While
the significance of the prefactor to the integral is not clear, the above
equation remains qualitatively accurate over a wide range near the Anderson
transition. Setting $\omega =i/\tau _{\varphi }$ and doing the $Q$-integral
in 3d, 
\bea
\frac{D_{0}}{D} &\approx & 1+\frac{1}{\pi mk_{F}}\int_{1/L_{\phi }}^{1/l}dQ\frac{%
Q^{2}}{DQ^{2}}\cr
&=& 1+\frac{D_{0}}{D}\frac{3}{\pi k_{F}^{2}l^{2}}-\delta
\left( \frac{D_{0}}{D}\right) ,
\label{delta}
\eea%
where 
\bea
\delta \equiv \frac{D_{0}}{D}\frac{3}{\pi k_{F}^{2}l^{2}}\frac{l}{%
L_{\varphi }}
\eea 
is assumed to be a small correction, and Eq.~(\ref{delta})
should not be solved self-consistently. This follows from the fact that the
diffusion coefficient of electrons at fixed energy entering the Cooperon
expression is that of non-interacting electrons, and is given by the limit $%
T\rightarrow 0$, $L_{\varphi }\rightarrow \infty $ and therefore $\delta
\rightarrow 0$. Then the correction at finite $T$ is given by 
\bea
\frac{D}{D_{0}} &=& \frac{1}{\left( \frac{D_{0}}{D}\right) _{0}-\delta \left( 
\frac{D_{0}}{D}\right) }\cr
&\approx & \left( \frac{D}{D_{0}}\right) _{0}+\left( \frac{D}{D_{0}}\right) _{0}
\frac{3}{\pi k_{F}^{2}l^{2}}\frac{l}{L_{\varphi }}%
\eea%
where 
\be
\lim_{T\rightarrow 0}\frac{D}{D_{0}}\equiv \left( \frac{D}{D_{0}}\right)
_{0}.
\ee%
Using the relation $\sigma _{3d}=e^{2} N_{0} D$ where the longitudinal
sheet conductance $\sigma _{\square }=\sigma _{3d}t$, with $t$ being the
film thickness, we finally get the temperature dependent weak localization
correction term 
\bea
\frac{\delta \sigma _{\square }}{L_{00}} &=& \left( \frac{D}{D_{0}}\right) _{0}%
\frac{2}{\pi }\frac{t}{L_{\varphi }}\cr
\left( \frac{D}{D_{0}}\right)_{0} &\approx &\frac{2}{1+\sqrt{1+4R_{0}^{2}/{a^{2}}}}
\label{WL}
\eea%
where $R_{0}=L_{00}/\sigma _{\square }(T$=$0)$, $L_{00}=e^{2}/\pi h$, $%
a=3\pi/2k_{F}tb_{0}$, $b_{0}$ is a number of order unity and we
have solved the self-consistent equation for $D$ in order to express $D_{0%
\text{ }}$in terms of $D$ and finally $R_{0}$. Thus in this case, the weak
localization correction has a prefactor which is not universal. While this
reduces to the well-known universal result at weak disorder $R_{0}\ll a$, it
becomes dependent on disorder characterized by the sheet resistance $R_{0}$
at strong disorder and at the same time substantially extends the 3d regime
near the transition. 

Using the expression for $L_{\varphi }$ (Eq.~(\ref{L-phi-3d})) into Eq.~(\ref%
{WL}), we finally obtain the total conductivity, including the quantum
correction to the conductivity due to weak localization in 3d arising from
scattering of electrons off antiferromagnetic spin waves in Mn, 
\begin{equation}
\frac{\sigma _{\square }}{L_{00}}=A+\frac{B}{\sqrt{\sinh [\Delta /T]}},
\label{sigmaWL}
\end{equation}%
\textbf{\textbf{}}where the parameter $A$ is temperature independent and the parameter 
\bea
B &\equiv & \left( \frac{D}{D_{0}}\right) _{0}\frac{2}
{\pi}\left( \frac{12\pi }{5}\right) ^{1/2}\left( \frac{\Delta }{\bar{J}}\right)^{3/2}tk_{F}\cr%
&=&\frac{2c}{1+\sqrt{1+4R_{0}^{2}/{a^{2}}}},
\label{BFit}
\eea%
where 
\be
c\equiv \left( \frac{\Delta }{\bar{J}}\right) ^{3/2}
\left( \frac{48t^{2}k_{F}^{2}}{5\pi} \right) ^{1/2}.
\label{cFit}
\ee%

The data presented here is for a single film prepared with an initial $R_0
\approx$~6~k$\Omega$. Disorder was consequently increased in incremental
stages up to 180~k$\Omega$ by annealing at approximately 280~K~\cite%
{misra_2011}. Additional samples were grown at intermediate disorder and
measured to check reproducibility.

\begin{figure}[tbp]
\begin{center}
\includegraphics[width=9cm]{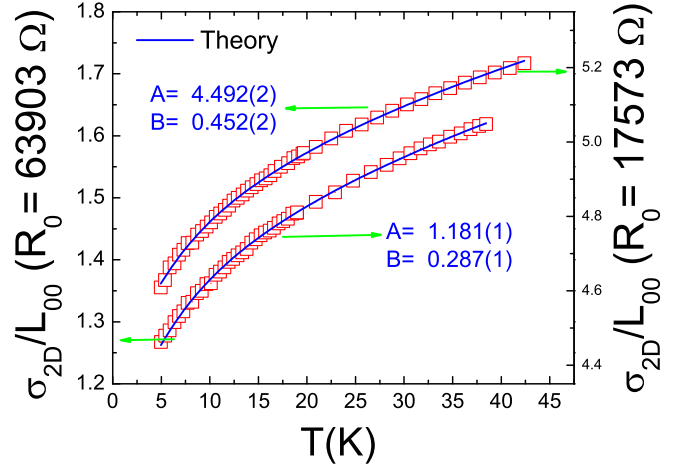}
\end{center}
\caption{The temperature-dependent normalized conductivity (open squares)
for two samples with the indicated disorder strengths of $R_0 =$~17573~$%
\Omega$ and 63903~$\Omega$ show good agreement with theory (solid lines).
The fitting parameters $A$ and $B$ are indicated for each curve with the
error in the least significant digit indicated in parentheses.}
\label{fig:cond}
\end{figure}

Figure~\ref{fig:cond} shows the conductivity data for two samples with
disorder $R_{0}=$~17573~$\Omega $ and 63903~$\Omega $ with corresponding
fittings to the expression (\ref{sigmaWL}) where $A$ and $B$ are taken as
fitting parameters and $\Delta =$~16~K is the spin wave gap. The fits are
sensitive to the parameters $A$ and $B$ but relatively insensitive to $%
\Delta $. We find that $\Delta =$~16~$\pm $~4~K provides good fittings in
the whole range of disorder (from 6 to 180~k$\Omega $).

\begin{figure}[tbp]
\begin{center}
\includegraphics[width=9cm]{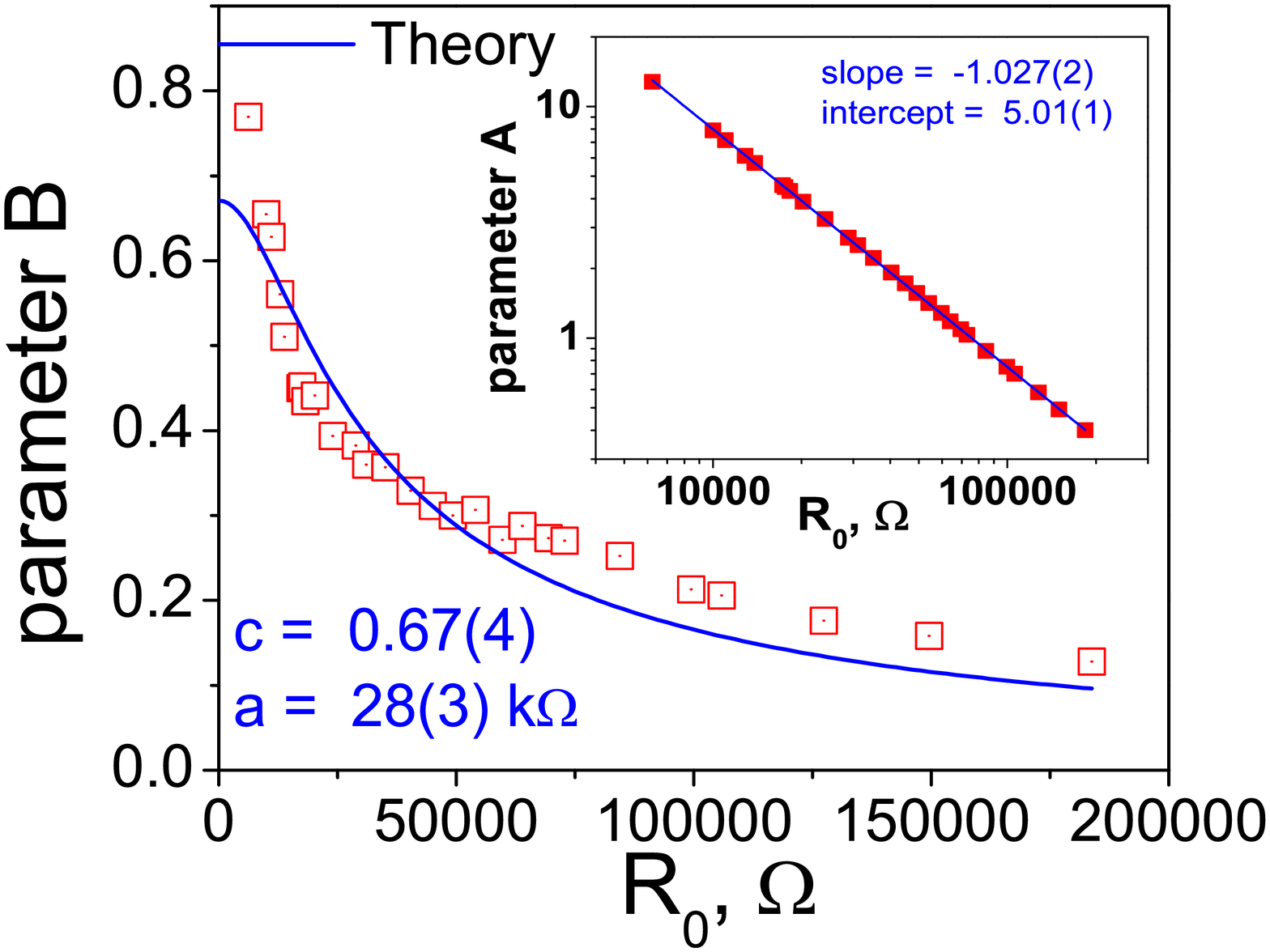}
\end{center}
\caption{Dependence of the fitting parameters $B$ and $A$ (inset) on
disorder $R_0$ for $\Delta=$~16~K. The fitting parameters are indicated for
each curve with the error in the least significant digit indicated in
parentheses.}
\label{fig:parb}
\end{figure}

Figure~\ref{fig:parb} shows the dependence of the parameter $B$ on the 
disorder strength $R_0$ (open squares) and a theoretical fit (solid line)
using Eq.~(\ref{BFit}), where $c$ and $a$ are fitting parameters. The solid 
line for this two-paramener fit is drawn for the best-fit values $c=0.67 \pm 
0.04$ and $a= 28 \pm 3$~k$\Omega$. We note that the fit is of reasonable
quality over most of the disorder range except for the film with the least 
disorder ($R_0 = 6$~k$\Omega$) where $B = 0.77$,
somewhat below the saturated value
$B = c = 0.67$ evaluated from Eq.~(\ref{BFit}) at $R_0 = 0$. Using higher 
values of $c$ (e.g., $c=0.8$) and lower values of $a$ (eg., $a = 22$~k$\Omega$)
improves the fit at low disorder strengths but 
increases the discrepancy at higher disorder strengths. 




Substituting the Fermi energy for bulk Mn~\cite{ashcroft_1976},
a thickness $t=2$~nm known to 20\% accuracy, together with the best-fit
value for $c$ into Eq.~(\ref{cFit}), we calculate the value $\bar{J} =$~320~$%
\pm$~93~K. Gao et al.~\cite{gao_2008} performed inelastic scanning tunneling
spectroscopy (ISTS) on thin Mn films and reported $\Delta$ in the range from
25 to 55~K and $\bar{J}=A_{s} k_F=$~3150~$\pm$~200~K. The agreement of energy 
gaps is acceptable; however our significantly lower value of $\bar{J}$ is 
possibly due to the high disorder in our ultra thin films. Also our model decription may be too simple to provide a quantitative description of all aspects.

Since the temperature-dependent correction $B/\sqrt{\sinh (\Delta /T)}$ of
Eq.~\ref{sigmaWL} is small compared to the parameter $A$, we can write
$\sigma_{\square} \approx 1/R_0$ so that Eq.~\ref{sigmaWL} reduces to the 
expression $A \approx 1/L_{00}R_0$. The logarithmic plot derived by taking the
logarithm of both sides of this approximation is shown in the inset of
Fig.~\ref{fig:parb}. The slope of -1 confirms the linear dependence of $A$ on
$1/R_0$ and the intercept of 5.01 (10$^{5.01}\approx $~102~k$\Omega$) is
within 20\% of the expected theoretical value 
$L_{00}= e^2/\pi h =$~81~k$\Omega $,
for the normalization constant. Accordingly, the conductivity corrections in
Eq.~\ref{sigmaWL} are small compared to the zero temperature conductivity and
the normalization constant $L_{00}$ for the conductivity is close to the
expected theoretical value.

Using Eq.~(\ref{WL}) and the obtained value for
$a\approx $~28~k$\Omega $ we can
compare the dephasing length ($L_{\varphi }$) with the thickness ($t\approx $%
~2~nm) at 16~K. For the sample with $R_{0}=$~63903~$\Omega $ the ratio $%
L_{\varphi }/t\approx $~0.5 and for the sample with $R_{0}=$~17573~$\Omega $, 
$L_{\varphi }/t\approx $~2. The latter estimate assumes no spin
polarization, while a full polarization would imply $L_{\varphi }/t\approx $%
~1. Thus $L_{\varphi }$ is smaller than or close to the thickness of the
film, which keeps the film in the three-dimensional regime for almost all
temperatures and disorder strengths considered.

In conclusion, we have performed \textit{in situ} transport measurements on
ultra thin Mn films, systematically varying the disorder ($R_{0}=R_{xx}$($T=$%
~5~K)). The obtained data were analyzed within a weak localization theory in
3d generalized to strong disorder. In the temperature range considered
inelastic scattering off antiferromagnetic spin waves is found to be strong giving rise to a
dephasing length shorter than the film thickness, which places these systems
into the 3d regime. The obtained value for the spin wave gap was close to
the one measured by Gao et al.~\cite{gao_2008} using ISTS, while the
exchange energy was much smaller.

This work has been supported by the NSF under Grant No 1305783 (AFH).
PW thanks A.\ M.\ \ Finkel'stein for useful discussions and acknowledges
partial support through the DFG research unit "Quantum phase transitions".

\bibliographystyle{apsrev}

\end{document}